# Material Decomposed Cargo Imaging with Dual Energy Megavoltage Radiography

Polad M. Shikhaliev [a)]

## Abstract

Radiography imaging has well-known applications in medicine, industry, security, and scientific research. Megavoltage (MV) radiography has important applications in imaging large cargos for detecting illicit materials. A useful feature of MV radiography is the possibility of decomposing and quantifying materials with different atomic numbers. This can be achieved by imaging cargo at two different x-ray energies, or dual energy (DE) radiography. The performance of both single energy and DE radiography depends on beam energy, beam filtration, radiation dose, object size, and object content. The purpose of this work was to perform comprehensive qualitative and quantitative investigations of the image quality in MV radiography depending on the above parameters. A digital phantom was designed including Fe background with thicknesses of 2cm, 6cm, and 18cm, and materials samples of Polyethylene, Fe, Pb, and U. The single energy images were generated at x-ray beam energies 3.5MV, 6MV, and 9MV. The DE material decomposed images were generated using interlaced low and high energy beams 3.5/6MV and 6/9MV. The x-rays beams were filtered by low-Z (Polyethylene) and high-Z (Pb) filters with variable thicknesses. The radiation output of the accelerator was kept constant for all beam energies. The image quality metrics was signal-to-noise ratio (SNR) of the particular sample over a particular background. It was found that the SNR depends on the above parameters in a complex way, but can be optimized by selecting a particular set of parameters. For some imaging setups increased filter thicknesses, while strongly absorbed the beams, increased the SNR of material decomposed images. Beam hardening due to polyenergetic x-ray spectra resulted in material decomposition errors, but this could be addressed using region of interest decomposition. It was shown that it is not feasible to separate the materials with close atomic numbers using the DE method. Particularly, Pb and U were difficult to decompose, at least at the dose levels allowed by radiation source and safety requirements.

[a)] Corresponding author:
4949 Stumberg Ln
Baton Rouge, LA 70816
pshikhal@yahoo.com

## 1. Introduction

Projection x-ray imaging (radiography) has well-known applications in medicine, industry, security, and scientific research. In each of these areas certain requirements exist that determine the data acquisition methods, detection systems, x-ray energies, radiation dose, and imaging time. One important application of projection radiography is imaging cargos with large sizes for detecting illicit materials such as explosives, narcotics, nuclear materials, high density radiation shields, etc. These cargos may have sizes of up to a few meters in cross sections and average densities of about thousand kilograms per cubic meter. To image such a large object high energy x-ray beams with high penetration capabilities are required. X-rays with mega-electronvolt energies are appropriate for this purpose, and the corresponding imaging technique is called mega-voltage (MV) radiography.

Another useful feature of projection x-ray imaging is the possibility of separating and quantifying materials with different effective atomic numbers. In order to separate two materials from each other, two x-ray images of the object should be acquired at two different energies in a single image acquisition. This imaging method is called dual energy (DE) radiography. As in the case of single energy radiography, DE radiography also has various applications such as bone density and contrast measurements in medical imaging, and airport baggage inspection in security imaging. DE MV (DEMV) radiography has been investigated for cargo imaging [1-8] and several systems are now commercially available [9-13].

There are several conceptually different methods for acquiring DE radiography data. These include using (1) x-ray source and detector where the x-ray pulses are dynamically switched (interlaced) between two energies, (2) x-ray source and a two-layer detector where the detector layers can separate low and high energy x-ray photons, and (3) x-ray source with an energy-resolved photon counting detector that can split the x-ray spectrum into several parts and acquire multi-energy x-ray images in a single acquisition. Each of these methods has its advantages and limitations that determine its feasibility for cargo radiography. For DEMV cargo radiography, the method using a source with interlaced DE x-ray pulses has been shown as feasible [3, 4, 6, 8], and corresponding x-ray sources are commercially available [12, 14].

There are several difficulties when DEMV cargo radiography is acquired using x-ray beams with interlaced energies. One limitation is that the low and high energy x-rays are filtered by the same filter because both rays are emitted from the same source. Such filtration may not be optimal for separation of the energies of two beams. This problem is further complicated because the x-ray attenuations of the high-Z (atomic number) materials do not change monotonically as opposed to low-Z materials due to higher pair production cross section. The image quality of single energy radiography and material selective DE materials, background materials, and materials to be imaged. In this work, we have performed comprehensive simulation studies of single energy and DEMV radiography using (1)different beam filter materials including low- and high-Z materials, (2)different background thicknesses, (3)different material samples for imaging, (4)single energy radiography with different energy settings, and (5) DE radiography with different energy settings. The signal to noise ratio (SNR) of the images was used as image quality metrics while the radiation output of the source was kept constant for all image acquisitions. The complex nature of the dependence of SNR on beam energies, sample types, filter types, and background thicknesses was demonstrated quantitatively and qualitatively. In general, high-Z and low-Z filters materials showed similar performances, while the higher beam energies were more preferable, particularly for larger background thicknesses. Interestingly, for some imaging tasks, increasing filter thickness improved SNR in DE material radiography depends on the types and thicknesses of the filter decomposed images despite the fact that the thicker filter absorbed more x-rays and decreased photon statistics. Feasibility of decomposing high-Z materials such as Pb and U by DE radiography was also tested.

## 2. Methods and materials

### 2.1. Imaging Phantom

As noticed above, the cargos may have various sizes, material contents, area densities, and spatial distributions of the included materials. Therefore, it is difficult to identify a single phantom setup for performing comprehensive quantitative and qualitative studies of DEMV cargo radiography. We designed a digital phantom which meets many of the representative parameters of real objects including sizes, materials of interests, and area densities, as shown in the **figure 1**.

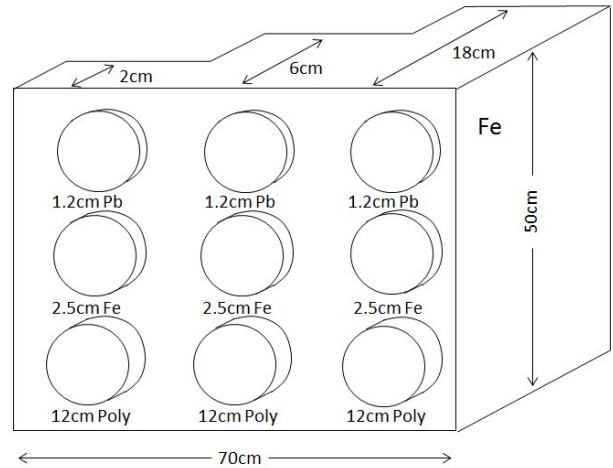

**Figure 1.** Digital Phantom used in the simulation study. It includes Fe background with thicknesses of 2cm, 6cm, and 18cm, respectively. The sample materials, including Pb, Fe, and Polyethylene with thicknesses of 1.2cm, 2.5cm, and 12cm, respectively, were placed at the top of Fe background with different thicknesses.

The phantom includes a Fe background material with a 50x70cm rectangular cross section and stepwise changing thicknesses of 2cm, 6cm, and 18cm. The samples of the three materials, including Pb, Fe, and Polyethylene with thicknesses of 1.2cm, 2.5cm, and 12cm, respectively, were placed over the areas of the background material with different thicknesses. The sample thicknesses were chosen such that they have similar area densities. Fe was chosen as the background material because it is one of the most widely used industrial materials, including also the material of the cargo container walls, and can represent an appropriate background material. Its thickness changes from minimal 2cm, which simulates cargo wall thickness, to 18cm, which is similar to higher area densities of the cargos. The sample materials Polyethylene, Fe, and Pb represent low-Z, medium-Z, and high-Z materials, respectively, covering materials of interest to be imaged in cargo radiography. Thus, the designed phantom covers probable background material type and thickness ranges as well as the range of the material samples to be imaged.

In one simulation, we substituted the Polyethylene sample with uranium (U) with 0.72cm thickness to evaluate possibility of decomposing two high-Z materials Pb and U from each other.

### 2.2. X-ray beams

The x-ray energies used in cargo radiography should be sufficiently high to penetrate large cargos with sizes of approximately $2.3x2.3m^2$ in cross-section [15]. The contents of the cargos can be highly non-uniform in terms of material types and filled volume fractions. It is known that approximately 80% of cargos exhibit average area densities of up to $160g/cm^2$ which corresponds to approximately 20cm thickness of steel [3, 4, 6].

the x-ray beam parameters to allow sufficient penetration of the x-rays through the objects. In general, the system should be able to image the sample over the background when the area density of the sample is 20% of the area density of the background [16]. The radiation dose applied to the cargo should also be limited to 500mRem effective dose (which is equivalent to In general, imaging larger thicknesses of up to 40cm steel equivalent would be desirable. There are ANSI standards that determine 5mGy air kerma) because of possibilities of stowaways in the cargo [6]. The above requirements are met when the x-ray energies in the MeV ranges are used. However, the energies should not be higher than 10MeV to minimize photonuclear reactions and neutron productions.

Previous works on DEMV cargo radiography have used various x-ray energies. Specifically, the interlaced x-ray pulses at low/high energies were used at 3.5MV/6MV [17], 4MV/8MV [3], and 6MV/9MV [5]. The above works used the fixed x-ray beam filters fabricated typically from high-Z materials such as W or Pb.

In the current study, we used beam energies of 3.5MV, 6MV, and 9MV for singe-energy radiography, and pairs of 3.5MV/6MV and 6MV/9MV for DEMV radiography. The x-ray beams were filtered with two types of filters, low-Z (Polyethylene) and high-Z (Pb) filters because low-Z and high-Z filters modify the x-ray spectrum differently which affects the DE material decomposition. For low-Z filters, Polyethylene with 6cm, 12cm, 24cm, and 48cm thicknesses were used. For high-Z filters, Pb with 0.5cm, 1cm, 2cm, and 4cm thicknesses were used. The thicknesses of the filters were selected such that respective low-Z and high-Z filters have similar area densities. **Figure 2** shows the filtered x-ray spectra at beam energies of 3.5MV, 6MV, and 9MV. For easy comparison of the filtered spectra, the numbers of the x-ray photons under each spectrum were normalized to one.

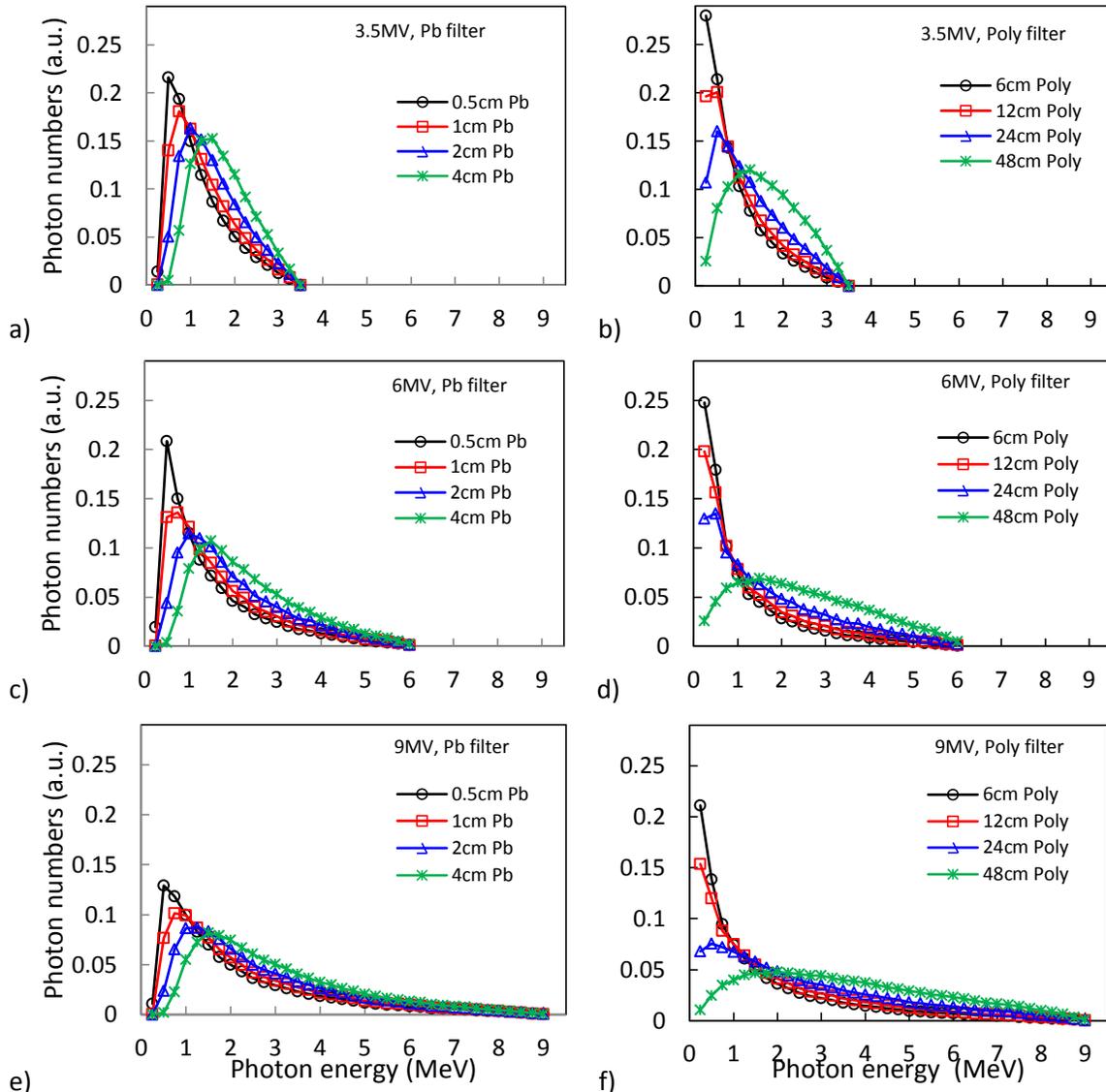

**Figure 2.** The photon energy spectra of the 3.5MV (a, b), 6MV (c, d), and 9MV (e, f) beams. The spectra were filtered by Pb filters (a, c, e) with 0.5, 1, 2, 4cm thicknesses, and Polyethylene filters (b, d, f) with 6, 12, 24, 48cm thicknesses, respectively. The filter thicknesses were selected such that corresponding area thicknesses of Pb and Polyethylene are the same.

**Figure 3** shows how the average photon energies differ when the beam is filtered with low-Z and high-Z filters with the same area densities. Although the thicker filters attenuate the x-ray beams more strongly and decrease the SNR in single-energy radiography images, this may have benefits in certain DE radiography images because filtration may help with better separation of the low and high energy spectra.

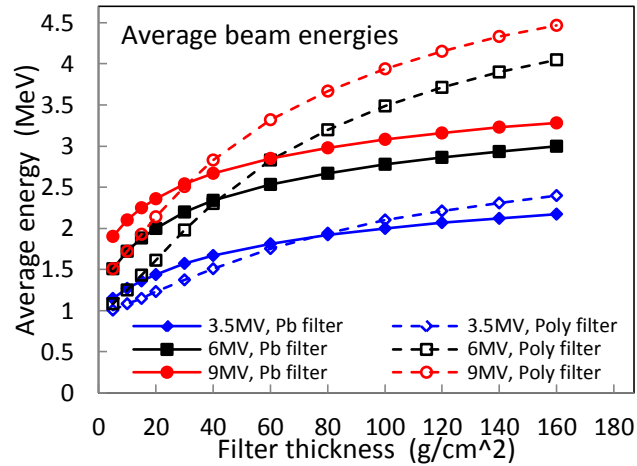

**Figure 3**. Average beam energies versus filter thicknesses. The filters with low and high atomic numbers change the average energies differently which can help to improve material decomposition in some imaging tasks.

**Figure 4** compares beam attenuations of low-Z and high-Z filters with the same area thicknesses. The low-Z filter provides slightly less beam attenuation than high-Z filter at the same area density.

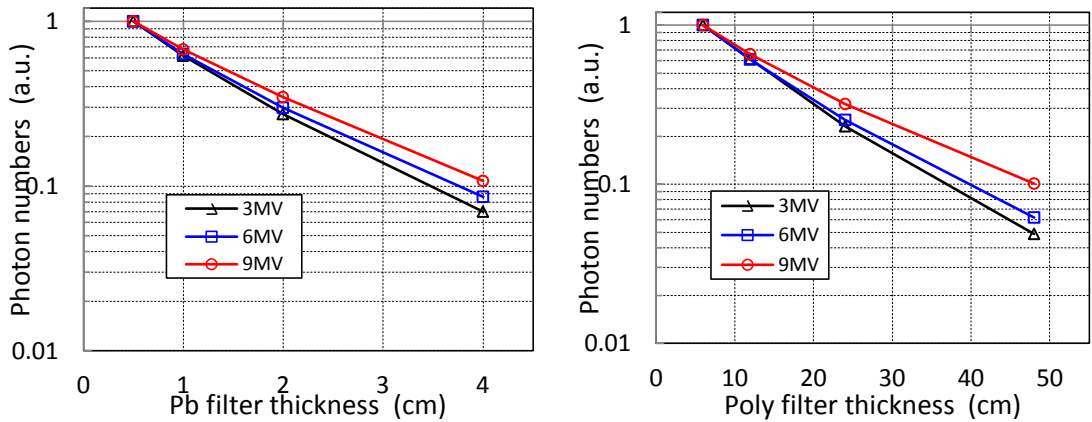

**Figure 4**. Photon attenuation factors versus filter thicknesses. The low-Z filter provides slightly less beam attenuation than high-Z filter at the same area density.

The radiation dose delivered by LINAC systems with interlaced beam energies is shared between the low and high energy pulses in certain proportion [3, 14]. For the commercially available system, Linatron-Mi, this proportion is 1:3 where the lower dose is delivered at lower energies [14]. We used the above dose sharing factor in our study. The Linatron-Mi system with interlaced low/high energy beams can provide maximum dose rates of 10/30Gy/min at 6/9MV, and 2.5/8Gy/min at 3.5/6MV, measured at 1m distance from the LINAC target [14]. In our study we used the total dose rate of 10.5Gy/min for both 3.5/6MV and 6/9MV settings. Taking into account the x-ray beam geometry and cargo scanning speed, the above dose rate provides the total dose (air kerma) of 0.875mGy which is well below the dose (air kerma) limit of 5mGy imposed in cargo scanning.

## 2.3. Theory of the material decomposition

The physical principle of the material decomposition in x-ray imaging is based on differences of changing of the linear attenuation coefficients of materials with x-ray energies [18-23]. In general, an object including $n$ known materials with different effective atomic numbers can be decomposed into components using $n$ x-ray images acquired at different energies. The term "decomposition" here means determination of the unknown thicknesses of each of the $n$ known materials included in the object. For simplicity, we assume that the object consists of two materials and is imaged at two x-ray energies that we call low and high energies $E_L$ and $E_H$, respectively. The linear attenuation coefficients of the first component at low and high energies are $\mu_{1L}$ and $\mu_{1H}$, respectively. Corresponding linear attenuation coefficients of the second component are $\mu_{2L}$ and $\mu_{2H}$, respectively. We further assume that the x-ray beam includes $N_{0L}$ and $N_{0H}$ photons with energies $E_L$ and $E_H$, respectively, and the thicknesses of the two components are $t_1$ and $t_2$, respectively. The low and high energy photon numbers in the beam passed through the object are $N_L = N_{0L}e^{-\mu_{1L}t_1 - \mu_{2L}t_2}$ and $N_H = N_{0H}e^{-\mu_{1H}t_1 - \mu_{2H}t_2}$, respectively. After log processing, the above expressions yield a system of linear equations with unknowns $t_1$ and $t_2$:

$$\begin{cases} \mu_{1L}t_1 + \mu_{2L}t_2 = \ln\dfrac{N_{0L}}{N_L} \\ \mu_{1H}t_1 + \mu_{2H}t_2 = \ln\dfrac{N_{0H}}{N_H} \end{cases} \quad (1)$$

The above system can be resolved with respect to thicknesses $t_1$ and $t_2$ using the known linear attenuation coefficients and photon numbers. After cancelling out the thickness $t_2$ in the above system the expression for the thickness $t_1$ is:

$$t_1 = \frac{\mu_{2H}\ln\dfrac{N_{0L}}{N_L} - \mu_{2L}\ln\dfrac{N_{0H}}{N_H}}{\mu_{1L}\mu_{2H} - \mu_{2L}\mu_{1H}} \quad (2)$$

The expression for the thickness $t_2$ can be found similarly. Also, the system of equations (1) can be extended to larger numbers of materials assuming that the images are acquired at the same numbers of energies. As can be seen from the expression (2), the thickness $t_1$ is undetermined if the condition

$$\frac{\mu_{1L}}{\mu_{2L}} = \frac{\mu_{1H}}{\mu_{2H}} \quad (3)$$

holds, which makes the denominator of (2) zero. Because the low and high energies $E_L$ and $E_H$ are independent, the expression (3) is equivalent to

$$\mu_1(E) = K\,x\,\mu_2(E) \quad (4)$$

for all energies $E$, where $K$ is a constant. The expression (4) holds if the two materials have the same effective atomic numbers and mass attenuation coefficients. Therefore, two materials can be decomposed using DE method only if they have different effective atomic numbers and linearly independent mass attenuation coefficients.

**Figure 5a** shows the dependence of the mass attenuation coefficients of several materials on photon energies, including low-Z (Polyethylene, $H_2O$), medium-Z (Al, Fe, Cu), and high-Z (Pb, U) materials. In the energy range of interest, 1-9MeV, the mass-attenuation coefficients of low-Z and medium-Z materials monotonically decrease, while for high-Z materials such as Pb and U it has a minimum at around 3MV due to a higher pair production cross-section. **Figure 5b** shows the relative mass-attenuation coefficients $(\mu/\rho)_f / (\mu/\rho)_k$ of the two materials $f$ and $k$. Specifically, the relative mass-attenuation coefficients of Polyethylene/Fe, Al/Fe, Cu/Fe, W/Fe, Pb/Fe, $H_2O$/Polyethylene, and Pb/U are shown. As can be seen, for some material pairs, it is possible to select a particular low and high energy set from the 1-9MeV range such that the expression (3) holds and decomposition is not possible.

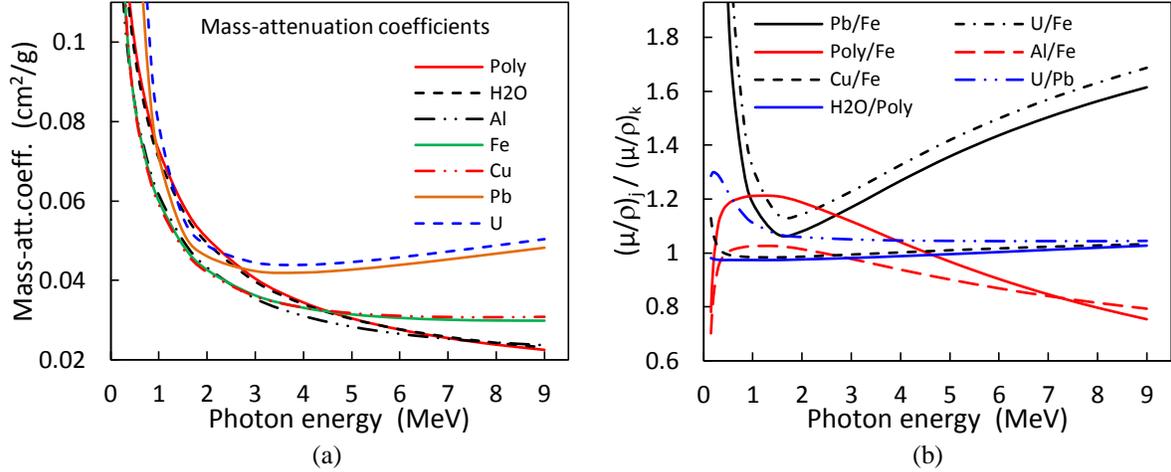

**Figure 5.** (a)The total mass-attenuation coefficients of seven materials with low, medium, and high atomic numbers plotted against photon energies. (b)Relative mass-attenuation coefficients of the materials. The two materials cannot be decomposed one from another by DE subtraction if the relative mass-attenuation coefficients for these two materials at low and high energy are equal.

For the material pairs with substantially different atomic numbers, it is possible to select low and high energy sets such that the expression (3) is far from equality, and decomposition is possible. However, for the material pairs with close atomic numbers such $H_2O$/Polyethylene and Pb/U, any low and high energy pair will result in approximate equality in the expression (3), and decomposition will not be particularly useful due to low SNR.

The above considerations and expressions (1)-(4) are correct for monoenergetic x-rays that are difficult to provide for cargo imaging applications, although some research in this area is ongoing [24]. For the polyenergetic x-rays used in practice, the x-ray beams are hardened as they pass through the materials. The average photon energies are increased and effective mass-attenuation coefficients are decreased depending on the material type and thicknesses. Therefore, the conditions (3)-(4) for the material decomposition become dependent on the thicknesses and types of the sample and background materials. In other words, when polyenergetic x-rays are used the coefficients in the system of equations (1) depend on background/sample thicknesses and types which may change for different areas of the object. Therefore, when polyenergetic x-rays are used, the decomposition should be performed separately for each particular area of interest of the object, which should not represent a problem when using dedicated computer software.

Notice that previous works on DE cargo imaging used material identification based on a different approach. Namely, the effective linear attenuation coefficients of the same material at low and high energies were determined and the relations of these two coefficients of the same material were tabulated and used to identify the unique material [3, 5, 8, 17]. However, if the two or more materials in the object do overlap, the identified material will be an unknown mixture with certain effective atomic number that may not correctly identify a particular material. The DE energy material decomposition method used in the current work allows for complete separation and quantification of two materials, with the potential of multi-material decomposition using multiple energies.

## 2.4. Simulation setup

The simulation setup included a megavoltage x-ray source with interlaced low/high energy pulses, the phantom, and a linear array of the detector pixels. The source-to-detector distance was 3m, and the source-to-phantom distance was 1.5m. The x-ray focal spot size was 2mm, as reported for the commercial system [14]. The pulse repetition frequency was 200pulse/s for the interlaced low and high energy pulses, and the pulse duration was 10-100µs [14]. The detector was a pixilated $CdWO_4$ scintillator connected to a linear array of photodiodes. The detector pixel size was 3x3mm² as viewed from the source side, and the detector thickness was 20mm [3, 25]. The length and arrangement of the linear array of the detector can be variable. For example, the detector array can be L-shaped to optimally cover the rectangular shape of the cargo. We used a linear shape with a detector length of 200cm that fully covered the phantom. The thickness of the fan beam was 3mm to match to the detector pixel size.

The speed of the cargo movement during the scanning process was assumed to be 60cm/s which is consistent with practical applications [3, 7]. With the above scan speed and at 200pulse/s pulse frequency, the sampling step of the object in the direction of the scan was 3mm, which matched the thickness of the fan beam.

For the monoenergetic radiation, the x-ray flux $N$ at the given energy $E$ and the dose (air kerma) rate $D$ are related as

$$D = kN\left(\frac{\mu(E)}{\rho}\right)_{air} E \qquad (5)$$

where $\left(\frac{\mu(E)}{\rho}\right)_{air}$ is the mass-energy absorption coefficient of the air, and $k$ is the known constant [26]. For the polyenergetic x-ray beam passed through the filter material with thickness $t$ the dose rate is determined as

$$D = \int_{E_1}^{E_2} k N(E) e^{-\mu(E)t} \left( \frac{\mu(E)}{\rho} \right)_{air} E dE \qquad (6)$$

where $N(E)$ is the photon energy spectrum and $E_1$ and $E_2$ are the lowest and highest photon energies, respectively. For the given total dose rate of 10.5Gy/min and dose distribution of 1:3 between the low and high energy beams, respectively, the dose rates at low and high energies were 2.9mGy/min and 7.9mGy/min. Using these dose rates, known x-ray spectral distribution, and minimal filter thicknesses of 0.5cm Pb and 6cm Polyethylene, the photon fluxes were calculated using the expression (6). From calculated photon flux, known x-ray pulse rate, sampling speed, and pixel size, the photon numbers which arrived at the detector pixel per sampling interval was calculated. The 2D distribution of the photon numbers $N_D^{ij}(E)$ with energy $E$ arrived at the detector pixels $(i, j)$ was built line-by-line in the scan direction using the detector data. The statistical (Poisson) noise was added to each photon number $N_D^{ij}(E)$. This photon distribution represented a radiography image acquired with an ideal photon counting detector.

The detectors used in the current MV radiography systems are scintillators connected to the photodiode array, that are energy integrating detectors. Their signal response to the arrived photons is determined as:

$$S_{ij} = \int_{E_1}^{E_2} N_D^{ij}(E) \left( 1 - e^{-\mu_D(E)t_D} \right) E \frac{\mu_D^{en}(E)}{\mu_D(E)} dE \qquad (7)$$

where $S_{ij}$ is the signal generated in the pixel $(i, j)$, $\mu_D(E)$ and $\mu_D^{en}(E)$ are the total and energy attenuation coefficients, respectively, and $t_D$ is the detector thickness. In the expression (7) the approximation was made that the secondary photons (Compton scattered and annihilation photons) leave the detector pixel after primary photon interaction while the generated particles (Compton electrons and electron-positron pairs) deposit their energies in the pixel [3]. The estimations show that this is generally a fair approximation taking into account that the pixel size 3x3x20mm$^3$ is small enough compared to the free path length of the high energy photons [27], and at the same time is large enough compared to secondary particles ranges [28].

## 2.5. Data processing

In cargo radiography, the single-energy images and DE material selective images are of interest. In this work, the single energy images were generated at the energies 3.5MV, 6MV, and 9MV, and the material decomposed images were generated using DE sets 3.5/6MV and 6/9MV. The SNR was used as a criterion for comparative image quality evaluations. Because low and high energy images are generated in a single image acquisition, it is desirable and possible to combine these two images to achieve an improved SNR. However, simple addition of these images would be suboptimal. The low and high energy images should be optimally weighted before being added together to take into account signal and noise characteristics of each image [29]. According to the generalized image weighting method, the highest SNR can be achieved if the $m$ independent images of the same object are weighted by the weighting factors

$$w_\gamma = \frac{A_\gamma}{\sigma_\gamma^2} \qquad (8)$$

and added, where $A_\gamma$ and $\sigma_\gamma$ are the signal and noise in the image $\gamma$, respectively [29]. This highest SNR is then determined as

$$SNR_{max} = \sqrt{\sum_{\gamma=1}^{\gamma=m} SNR_\gamma^2} \qquad (9)$$

where $SNR_\gamma$ is the signal-to-noise ratio of the image $\gamma$ [29]. It is important to note that because the weighting factors (8) depend on the signal and noise for the particular material sample, it is material-dependent. In other words, the weighting factors should be determined separately for each area of interest in the image, which can easily be performed using dedicated computer software.

The material decomposition was performed using the method described in the section 2.1, where the effective linear attenuation coefficients were derived directly from the log-processed low and high energy images.

## 3. Results

**Figure 6** shows the radiography images of the phantom acquired with 3.5/6MV and 6/9MV settings. To create a single radiography image from the 3.5/6MV image pair, the 3.5MV and 6MV images were optimally weighted and summed to maximize the SNR, and then log-processed to compress the dynamic range for better visualization. The 6/9MV images were processed similarly. The total dose rate was 10.5Gy/min for each of the settings 3.5/6MV and 6/9MV. The dose sharing between the low and high energy images were in a ratio of 1:3, respectively. The SNR in the Pb images over an 18cm Fe background, acquired at 3.5/6MV and 6/9MV with Pb and Polyethylene filters, are also shown in this figure. As can be seen, for general radiography imaging of 1.2cm Pb over an 18cm Fe background, the 6/9MV setup provides approximately 20% higher SNR than 3.5/6MV setup, while the two filter materials perform similarly for the different energy setups.

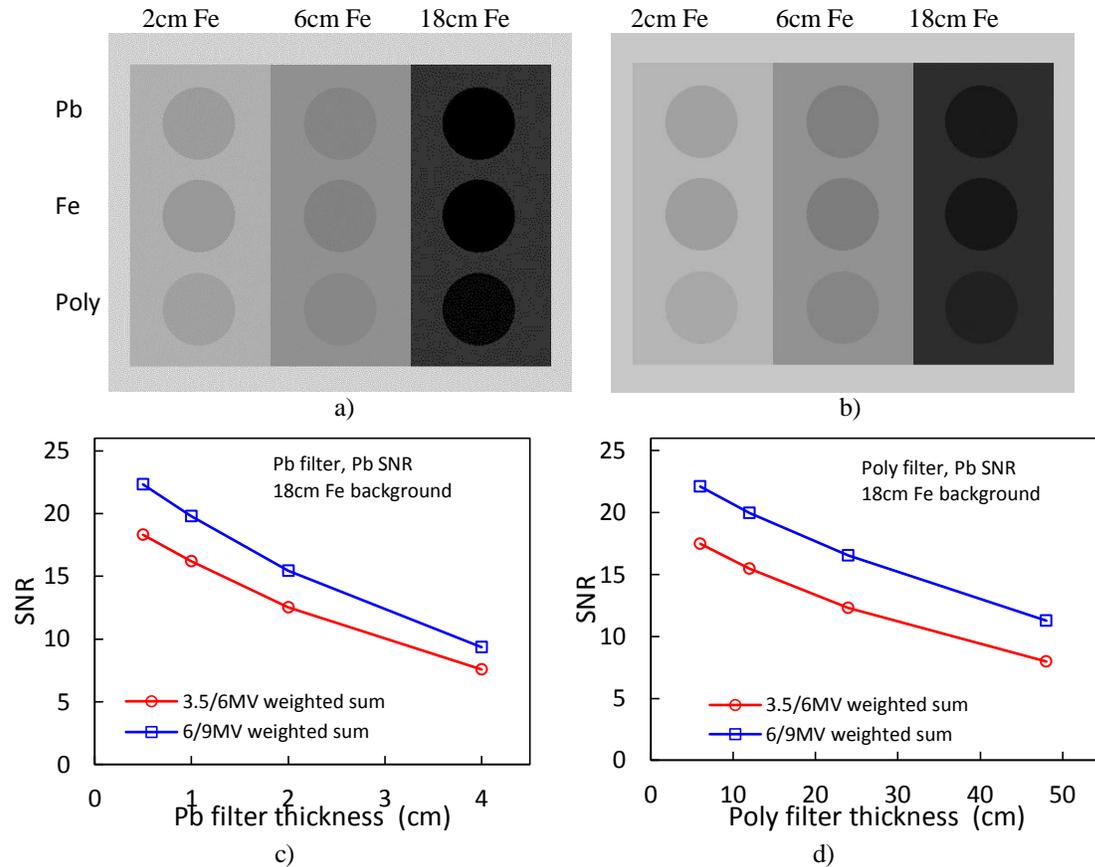

**Figure 6.** Radiography images (log processed) of the phantom after optimally weighting and summing low and high energy images (a) 3.5MV/6MV, and (b) 6MV/9MV. The total doses for 3.5/6MV and 6/9MV acquisitions were the same and equal 10.5Gy/min. The dose sharing between the low and high energy images were 1:3, respectively. The SNR of the Pb over 18cm Fe background, acquired at 3.5/6MV and 6/9MV with Pb (c) and Polyethylene (d) filters are shown.

Figure 7 shows the SNR of the Pb samples in single-energy MV radiography images plotted against LINAC energy. The Pb samples were placed at the top of the Fe backgrounds with 2cm, 6cm, and 18cm thicknesses. The beams were filtered with Pb filters with 0.5cm, 1cm, 2cm, and 4cm thicknesses, and with Polyethylene filters with 6cm, 12cm, 24cm, and 48cm thicknesses. For smaller background thicknesses the SNR decreases as LINAC energy increases. For larger background thicknesses the effect is opposite.

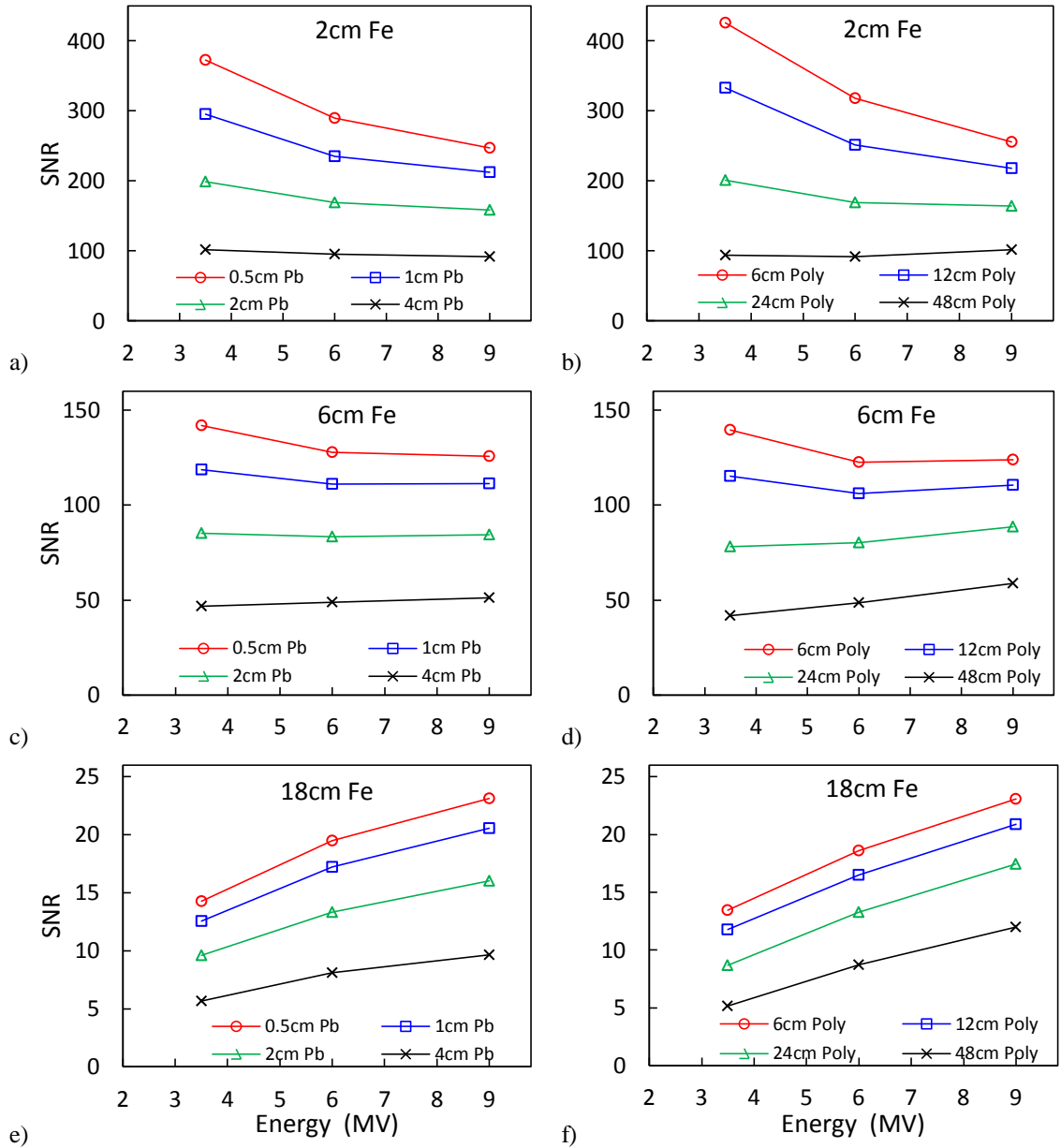

**Figure 7.** SNR of the Pb samples in single-energy MV radiography images plotted against LINAC energy. The Pb samples were placed at the top of the Fe background with 2cm (a,b), 6cm (c,d), and 18cm (e,f) thicknesses. The LINAC beams were filtered with Pb filters (a,c,e) with 0.5cm, 1cm, 2cm, and 4cm thicknesses, and with Polyethylene filters (b,d,f) with 6cm, 12cm, 24cm, and 48cm thicknesses.

**Figure 8** shows the DE subtracted images of the phantom using 3.5/6MV and 6/9MV energies. The x-ray beams were filtered with 0.5cm Pb. The contrast samples Pb, Fe, and Polyethylene were cancelled one sample at a time over 2cm, 6cm, 18cm Fe backgrounds. The contrast samples were cancelled for each Fe background thickness separately to avoid beam hardening artifacts and errors due to the polyenergetic nature of the used x-ray beams.

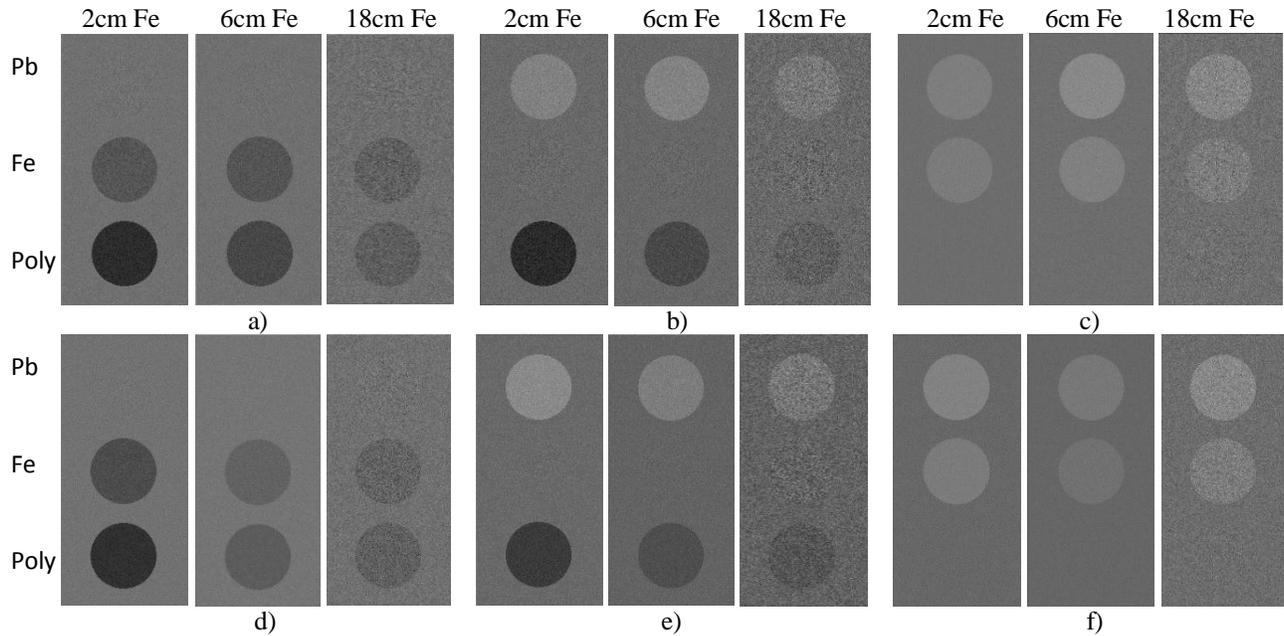

**Figure 8.** DE subtracted images of the phantom using low/high energy pairs of 3.5/6MV (a,b,c) and 6/9MV (d,e,f). The x-ray beams were filtered with 0.5cm Pb. The contrast samples Pb (a,d), Fe (b,e), and Polyethylene (c,f) were cancelled over the 2cm, 6cm, 18cm Fe background. Cancellation of the contrast samples were performed for each Fe background thickness separately to avoid beam hardening artifacts and errors due to the polyenergetic nature of the used x-ray beams.

**Figure 9** shows DE subtracted images of the phantom using low/high energy pairs of 3.5/6MV and 6/9MV. The x-ray beams were filtered with 0.5cm Pb, and the 2cm Fe background was cancelled. While the signals of the Fe contrast samples over 2cm Fe backgrounds are nearly zero, the residual signal from Fe samples remains over 6cm and 18cm Fe backgrounds. This is due to the beam hardening effects associated with the polyenergetic nature of the x-ray beams.

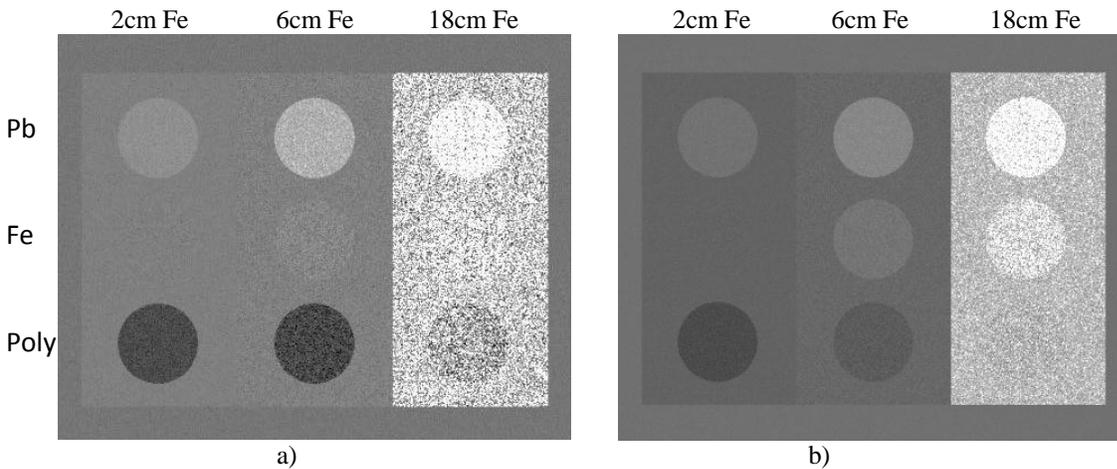

**Figure 9.** DE subtracted images of the phantom using low/high energy pairs of 3.5/6MV (a) and 6/9MV (b). The x-ray beams were filtered with 0.5cm Pb, and the 2cm Fe background was cancelled. While the signals of the Fe contrast samples over 2cm Fe backgrounds are nearly zero, the residual signal from Fe samples remains over 6cm and 18cm Fe backgrounds. This is due to the beam hardening effects associated with the polyenergetic nature of the x-ray beams.

**Figure 10** shows SNR of Pb samples after DE subtraction and cancellation of the Fe background, plotted against Pb and Polyethylene filter thicknesses. The energy sets 3.5/6MV and 6/9MV were used for DE acquisitions. The doses for the energy sets were the same. The SNR were determined at sample locations where Fe background thicknesses were 2, 6, and 18cm. As can be seen, increasing filter thickness increases SNR in some cases, despite the fact that the photon numbers and doses in the beam decrease.

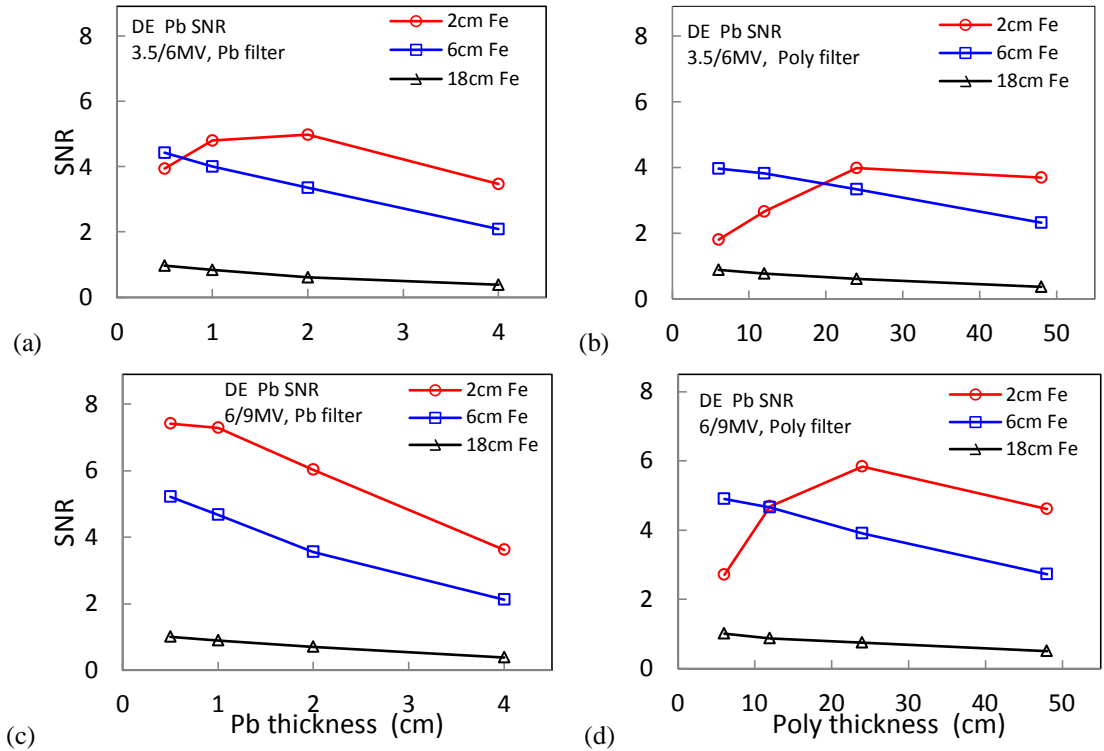

**Figure 10.** SNR of Pb samples after DE subtraction and cancellation of Fe background, plotted against Pb (a, c) and Polyethylene (b, d) filter thicknesses. The energy sets 3.5/6MV (a, b) and 6/9MV (c, d) were used for DE acquisitions at the same dose outputs of the sources. The SNR were determined at sample locations with Fe background thicknesses of 2, 6, and 18cm. Notice that increasing filter thickness increases SNR in some cases despite the fact that the photon numbers and dose in the beam decreases.

**Figure 11** shows SNR of Polyethylene samples after DE subtraction and cancellation of Fe background, plotted against Pb and Polyethylene filter thicknesses. The energy sets 3.5/6MV and 6/9MV were used for DE acquisitions. The doses for the energy sets were the same. The SNR were determined at sample locations where Fe background thicknesses were 2, 6, and 18cm. As seen, in some cases increasing filter thickness increases SNR despite the fact that the photon numbers and dose decrease.

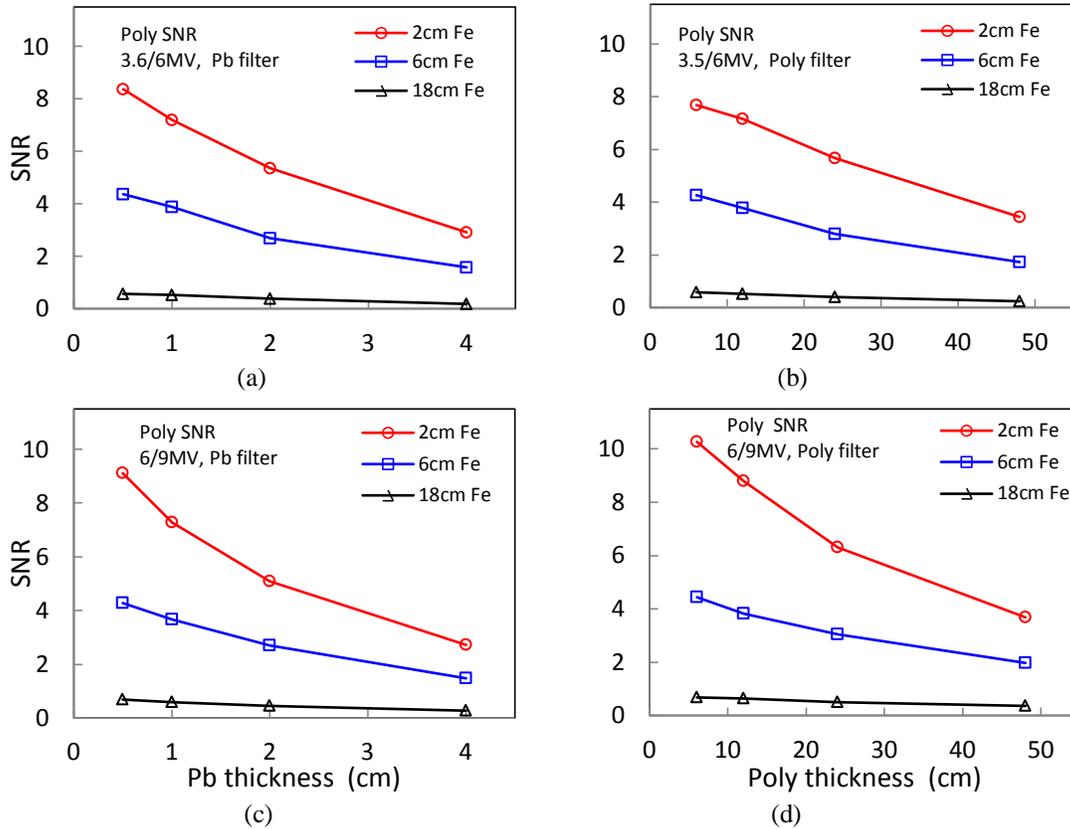

**Figure 11.** SNR of Polyethylene sample after DE subtraction and cancellation of the Fe background, plotted against Pb (a, c) and Polyethylene (b, d) filter thicknesses. The beam energy sets 3.5/6MV (a, b) and 6/9MV (c, d) were used for DE acquisitions at the same dose output of the sources. The SNR were determined at the sample locations where Fe background thicknesses were 2, 6, and 18cm.

**Figure 12** shows the decomposition factor (D-factor) determining the efficiency of Pb decomposition from the Fe background. When D-factor is 1, decomposition is not possible, and as the D-factor differs from 1, decomposition becomes more efficient with a higher SNR. The low/high energy sets 3.5/6MV and 6/9MV with Pb and Polyethylene filtrations were used. The Fe background thicknesses were 2cm, 6cm, and 18cm.

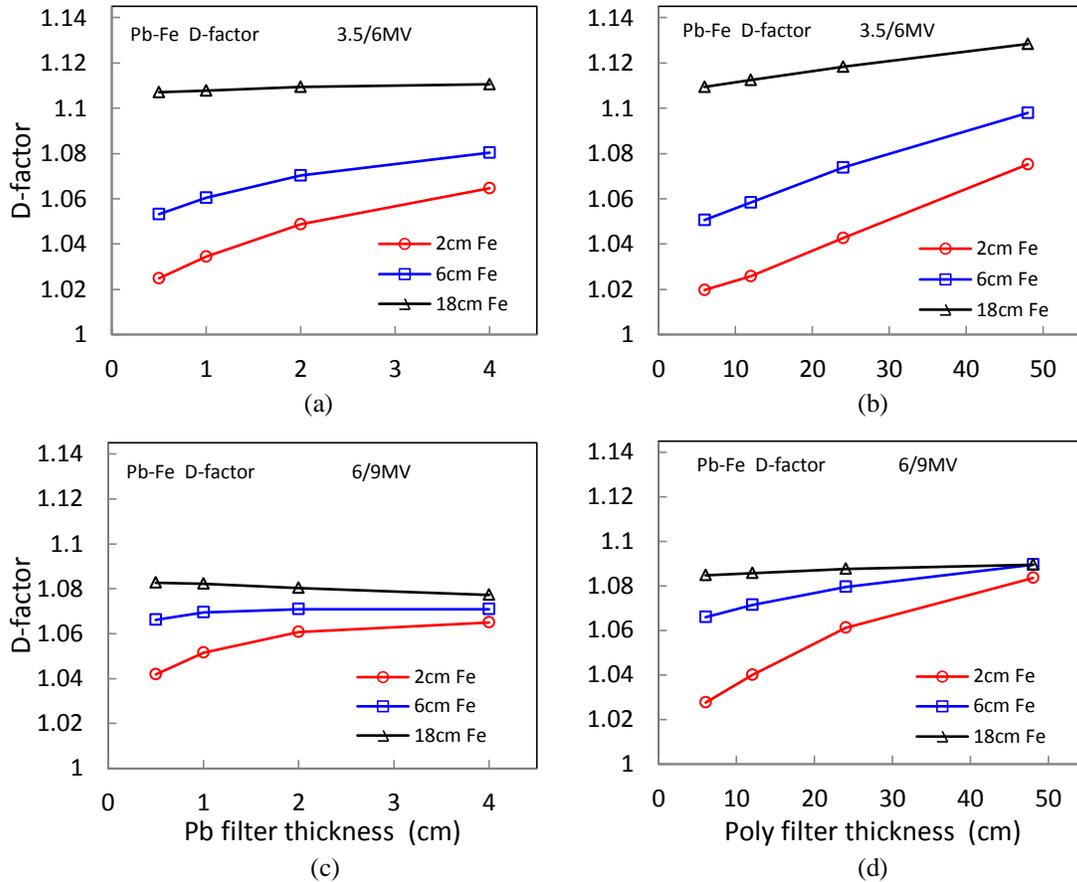

(a)

(b)

(c)

(d)

**Figure 12.** Decomposition factor (D-factor) determining the efficiency of Pb decomposition from the Fe background. When the D-factor is 1, decomposition is not possible, and as the D-factor differs from 1, decomposition becomes more efficient with higher SNR. The low/high energy sets 3.5/6MV (a,b) and 6/9MV (c,d) with Pb (a,c) and Polyethylene (b,d) filtrations were used. The Fe background thicknesses were 2cm, 6cm, and 18cm.

**Figure 13** shows the single energy radiography images acquired with 6MV and 9MV when 6/9MV energy set was used, and the DE subtracted image where Pb was cancelled out. In this phantom setup, the polyethylene sample was replaced by uranium (U) with 0.72cm thickness. As seen, the signal from U is nearly invisible when Pb is cancelled out. This indicates that it is difficult to decompose the two high-Z materials such as Pb and U using DE radiography because both materials are canceled out simultaneously.

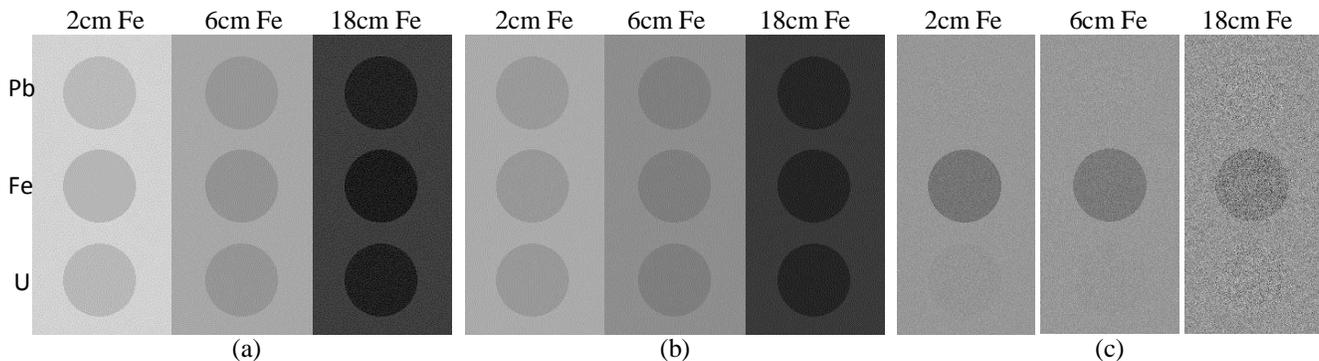

(a)

(b)

(c)

**Figure 13.** Radiography images of uranium (U) acquired with 6/9MV energy set. (a) 6MV and (b) 9MV single energy images, and (c) Pb cancelled DE subtracted image are shown. In this phantom setup the polyethylene sample was replaced by U with 0.72cm thickness. As seen, the signal from U is nearly invisible when Pb is cancelled out. This indicates that it is difficult to decompose two high-Z materials such as Pb and U using DE radiography as explained in the text.

## 4. Discussions and conclusion

We have reported the results of the comprehensive quantitative and qualitative simulation studies of cargo radiography. The study covered: (1)Single energy radiography with 3.5MV, 6MV, 9MV energies; (2)DE material decomposed radiography with 3.5/6MV and 6/9MV energy sets; (3)Beam filtration with low-Z and high-Z filters; (4)Imaged samples including low-Z, medium-Z, and high-Z materials; and (5)Variable background thicknesses. All studies were performed using interlaced DE beams with a fixed radiation output of the x-ray source.

The study showed that when interlaced DE beams are used, the low and high energy images can be optimally weighted and combined to provide a single image with the improved SNR. In single energy radiography with larger background thicknesses of 18cm Fe, the 6/9MV set provided an approximately 20% higher SNR than the 3.5/6MV set at the same radiation dose. The low-Z (Polyethylene) and high-Z (Pb) filters in 3.5/6MV and 6/9MV DE sets provided similar SNR as shown in **figure 6**. There was no particular beam energy that could be optimal for all single energy radiography setups as seen in **figure 7**. At the same radiation dose, the lower beam energies provided higher SNR for thinner material thicknesses, and vice versa. Because imaging thick materials with highest possible SNR is of interest, the 6/9MV set had an advantage over the 3.5/6MV set.

In DE imaging, any one material could be completely eliminated from the image as shown in **figure 8**. For an object including only two materials, any one material can be eliminated and remaining material can be quantified allowing two material decomposition and quantification using DE radiography. However, using polyenergetic x-rays created beam hardening problems that required region of interest material decomposition as shown in **figure 8**. In this case a separate decomposition was performed that was optimized for each of the Fe background thicknesses. If we cancelled out a particular material over only the 2cm Fe background, then this same material would not be fully eliminated over the 6cm and 18cm Fe backgrounds due to the beam hardening as shown in **figure 9**. Although this effect creates potential inaccuracy, properly designed DE software and calibration procedure can provide accurate region of interest dependent DE subtraction.

A useful effect was observed (**figure 10**) that for some imaging configurations, increasing Pb filter thickness from 0.5cm to 2cm, and Polyethylene filter thickness from 6cm to 24cm, substantially increased SNR of Pb samples. The SNR improvement was by a factor of 1.25 and 2 with Pb and Polyethylene filters, respectively. This SNR improvement occurred despite the fact that the increased filter thicknesses decreased beam intensity and dose by a factor of 3. The reason for this effect is better separation of the low and high energy spectra with the thicker filter. This effect can provide some benefits such as dose minimization at increased SNR, or substantial SNR improvement by increasing beam intensity at the dose levels still below the required limit.

Another useful observation is that it is nearly impossible to decompose materials with close atomic numbers. It was important to demonstrate that it is difficult to decompose two high-Z materials such as Pb and U. Cancelling out of a Pb sample resulted in near elimination of the U signal (**figure 13**). Theoretically, such decomposition could be possible if the radiation dose is greatly increased to provide a sufficiently high SNR of the U sample. However, the required x-ray outputs may not be available with current systems, and this may result in high object dose and shielding problems.

In this work we used the system parameters consistent with existing DEMV cargo radiography systems. However, there are several unused potentials that, if properly applied, could enhance capabilities of the DEMV radiography systems. One potential improvement could be the use of monoenergetic radiation that could eliminate beam hardening problems and provide better separation of beam energies. This possibility is being investigated by another group [24]. Another possibility is using polyenergetic x-ray sources and energy selective detectors. One possibility is using a dual-layer detector in which the first and second layers detect preferably lower and higher energy photons, respectively [30, 31]. Another approach is using photon counting spectroscopic detectors that can split the energy spectrum into several energy bins so that multi-energy data can be acquired in a single scan [32-36]. However, practical application of spectroscopic photon counting detectors in MV radiography could be difficult due to incomplete energy absorption, and high count rates. The authors [8] mentioned using non-spectroscopic energy sensitive detectors in MV cargo radiography, but no details were disclosed. Nevertheless, energy spectroscopy of the MV x-ray beams is extremely challenging task, and further efforts are needed to investigate this problem.

In conclusion, the DEMV cargo radiography was shown to be a multidimensional imaging task that depends on beam energies, filtrations, imaged samples, and object background. The results of this work demonstrated nature of these dependences both quantitatively and qualitatively, and possibilities of optimizations for particular imaging tasks.